\documentstyle[12pt]{article}
\begin{document}
\title{\bf $(q,h)$--analogue of Newton's binomial formula } 
\author{ H. B. Benaoum  \\  
Institut f\"ur Physik , Theoretische Elementarteilchenphysik, \\      
Johannes Gutenberg--Universit\"at, 55099 Mainz, Germany \\
email : benaoum@thep.physik.uni-mainz.de
} 
\date{20 Oktober 1998 \\
Revised 20 December } 
\maketitle
~\\
\abstract{In this letter, the $(q,h)$--analogue of Newton's binomial   
formula is obtained in the $(q,h)$--deformed quantum plane   
which reduces for $h=0$ to the $q$--analogue. 
For $(q=1,h=0)$, this is just the usual one as it should be. 
Moreover, the $h$--analogue is recovered for $q=1$.   
Some properties of the $(q,h)$--binomial coefficients are also given. \\
This result will contribute to an introduction of the $(q,h)$ 
--analogue of the well--known functions, $(q,h)$--special functions 
and $(q,h)$--deformed analysis.}
~\\
~\\
~\\
~\\
~\\
~\\
{\bf MZ--TH/98--44 }
\newpage
~\\
The $q$--analysis is an extension of the ordinary analysis, by the addition 
of an extra parameters $q$. When $q$ tends towards one the usual analysis is 
recovered. Such $q$--analysis has appeared in the literature long time 
ago~\cite{gas}. 
In particular, a $q$--analogue of the Newton's binomial formula, 
well--known functions like $q$--exponential, $q$--logarithm, $\cdots$ etc, 
the special functions arena's $q$--differentiation and $q$--integration 
have been introduced and studied intensively. \\
In~\cite{ben}, I have introduced the $h$--analogue of Newton's binomial  
formula leading thefore to a new analysis, called $h$--analysis. \\
In this letter, I will go a step further by generalizing the work~
\cite{ben}.
Indeed, an analogue of Newton's binomial formula is introduced here in 
the $(q,h)$--deformed quantum plane ( i.e. $(q,h)$ Newton's binomial 
formula which generalizes Sch\"utzenberger's formula~\cite{sch} with 
an extra parameter $h$. ) 
leading therefore to a more generalized analysis, called 
$(q,h)$--analysis. \\
With this generalization, the $q$--analysis, $h$--analysis and ordinary 
analysis are recovered respectively by taking $h=0$, $q=1$ and 
$(q=1,h=0)$. \\
~\\
Newton's binomial formula is defined as follows :
\begin{eqnarray}
( x + y )^n & = & \sum^n_{k=0} \left( \begin{array}{c}
n \\
k \end{array} \right) y^k x^{n-k}
\end{eqnarray}
where $\left( \begin{array}{c} 
n \\
k \end{array} \right) = \frac{n!}{k! (n-k)!}$  
and it is understood here that the 
coordinate variables $x$ and $y$ commute, i.e. $x y = y x$. \\
A $q$--analogue of (1), for the $q$--commuting coordinates $x$ and $y$ 
satisfying $x y = q y x$, first appeared in literature in 
Sch\"utzenberger~\cite{sch}, see also Cigler~\cite{cig},  
\begin{eqnarray}
( x + y )^n & = & \sum^n_{k=0} \left[ \begin{array}{c}
 n \\
 k  \end{array} \right]_q y^k x^{n-k}
\end{eqnarray}
where the $q$--binomial coefficient is given by :
\begin{eqnarray*}
\left[ \begin{array}{c} 
n \\
k \end{array} \right]_q & = & \frac{[n]_q!}{[k]_q! [n-k]_q!} 
\end{eqnarray*}
with 
\begin{eqnarray*}
[j]_q & = & \frac{1 - q^j}{1 - q} 
\end{eqnarray*}
~\\
The $h$--analogue has been introduced and defined in~\cite{ben} as follows : 
\begin{eqnarray}
( x + y )^n & = & \sum^n_{k=0} \left[ \begin{array}{c}
n \\
k \end{array} \right]_h y^k x^{n-k} 
\end{eqnarray}
provided that $x$ and $y$ satisfy to $[x,y] = h y^2$ and the $h$--
binomial coefficient $\left[ \begin{array}{c}
n \\
k \end{array} \right]_h$   
is given by :
\begin{eqnarray}
\left[ \begin{array}{c}
n \\
k \end{array} \right]_h & = & \left( \begin{array}{c} 
n \\
k \end{array} \right) h^k (h^{-1})_k
\end{eqnarray}
where $(a)_k = \Gamma(a+k)/\Gamma(a)$ is the shifted factorial. \\
~\\ 
Now consider Manin's $q$--plane $x' y' = q y' x'$.By the following linear 
transformation~( see \cite{mad} and references therein ) :
\begin{eqnarray*}
\left( \begin{array}{c} 
x' \\
y' \end{array} \right) & = & \left( \begin{array}{cc} 
1 & \frac{h}{q-1} \\
0 & 1 \end{array} \right) \left( \begin{array}{c} 
x \\
y \end{array} \right)
\end{eqnarray*}
Manin's $q$--plane changes to :
\begin{eqnarray}
x y & = & q y x + h y^2 
\end{eqnarray}
Even though the linear transformation is singular for $q=1$, the resulting 
quantum plane is well-defined. \\
~\\
{\bf \underline{Proposition~1 :} } \\
Let $x$ and $y$ be coordinate variables satisfying (5), then the following 
identities are true :
\begin{eqnarray}
x^k y & = & \sum^k_{r=0} \frac{ [k]_q!}{ [k-r]_q!} 
q^{k-r} h^r y^{r+1} x^{k-r} \nonumber \\ 
x y^k & = & q^k y^k x + h~ [k]_q~ y^{k+1}
\end{eqnarray}
These identities are easily proved by successive use of (5). \\
~\\
{\bf \underline{Proposition~2 :}} ( $(q,h)$--binomial formula ) \\
Let $x$ and $y$ be coordinate variables satisfying (5), then we have :
\begin{eqnarray}
( x + y )^n & = & \sum^n_{k=0}  \left[ \begin{array}{c} 
n \\
k \end{array} \right]_{(q,h)} y^k x^{n-k}
\end{eqnarray}
where $\left[ \begin{array}{c} 
n \\
k \end{array} \right]_{(q,h)}$ are the $(q,h)$--binomial 
coefficients given as follows :
\begin{eqnarray}
\left[ \begin{array}{c} 
n \\
k \end{array} \right]_{(q,h)} & = &  
\left[ \begin{array}{c}
n \\
k \end{array} \right]_q 
h^k (h^{-1})_{[k]} .
\end{eqnarray}
with $\left[ \begin{array}{c}
n \\
0 \end{array} \right]_{(q,h)} = 1$ and
\begin{eqnarray} 
(a)_{[k]} & = & \Pi_{j=0}^{k-1} ( a + [j]_q )
\end{eqnarray}
since by definition $[0]_q = 0$. \\   
~\\
Proof : \\
This proposition will be proved by recurrence. 
Indeed for $n=1,2$, it is verified. \\
Suppose now that the formula is true for $n-1$, which means :
\begin{eqnarray*}
( x + y ) ^{n-1} & = & \sum^{n-1}_{k=0} \left[ \begin{array}{c} 
n - 1 \\
k \end{array} \right]_{(q,h)} y^k x^{n-1-k},
\end{eqnarray*}
with $\left[ \begin{array}{c} 
n-1 \\
0 \end{array} \right]_{(q,h)} = 1$. \\
To show this for $n$, let first consider the following expansion :
\begin{eqnarray*}
( x + y )^n & = & \sum^n_{k=0} C_{n,k} y ^k x^{n-k} 
\end{eqnarray*}
where $C_{n,k}$ are coefficients depending on $q$ and $h$. \\
Then, we have : 
\begin{eqnarray*}
( x + y )^n & = & ( x + y ) ( x + y )^{n-1} \nonumber \\
& = & ( x + y ) \sum^{n-1}_{k=0} \left[ \begin{array}{c} 
n -1 \\
k \end{array} \right]_{(q,h)} y^k x^{n-1-k} \nonumber \\
& = & \sum^{n-1}_{k=0} \left[ \begin{array}{c} 
n -1 \\
k \end{array} \right]_{(q,h)} x y^k x^{n-1-k}~+
~ \sum^{n-1}_{k=0} \left[ 
\begin{array}{c} 
n -1 \\
k \end{array} \right]_{(q,h)} y^{k+1} x^{n-1-k}.
\end{eqnarray*}
Using the result of the first proposition, we obtain :
\begin{eqnarray*}
( x + y )^n & = & \left[ \begin{array}{c} 
n - 1 \\
0 \end{array} \right]_{(q,h)}~+~ 
\sum^{n-1}_{k=1} \left[ \begin{array}{c}
n - 1 \\
k \end{array} \right]_{(q,h)} q^k y^k x^{n-k}~+~ \nonumber \\
&  & \sum^{n-1}_{k=1} \left[ \begin{array}{c} 
n - 1 \\
k \end{array} \right]_{(q,h)} ( 1 + h~[k]_q ) y^{k+1} 
x^{n-1-k}~+~\left[ \begin{array}{c}
n - 1 \\
0 \end{array} \right]_{(q,h)} y x^{n-1} 
\end{eqnarray*}
which yields respectively :
\begin{eqnarray*}
C_{n,0} & = & \left[ \begin{array}{c} 
n - 1 \\
0 \end{array} \right]_{(q,h)}~=~1, \nonumber \\
C_{n,1} & = & q \left[ \begin{array}{c}
n - 1 \\
1 \end{array} \right]_{(q,h)}~+~\left[ \begin{array}{c} 
n - 1 \\
0 \end{array} \right]_{(q,h)}~=~\left[ \begin{array}{c} 
n \\
1 \end{array} \right]_{(q,h)} \nonumber \\
C_{n,k} & = & q^k \left[ \begin{array}{c} 
n - 1 \\
k \end{array} \right]_{(q,h)}~+
~( 1 + h~[k-1]_q ) \left[ \begin{array}{c} 
n - 1 \\
k - 1 \end{array} \right]_{(q,h)}~=~ \left[ \begin{array}{c} 
n \\
k \end{array} \right]_{(q,h)} ,  \nonumber \\
C_{n,n} & = & ( 1 + h~[n-1]_q ) ~\left[ \begin{array}{c} 
n - 1 \\
n - 1 \end{array} \right]_{(q,h)} ~=~
\left[ \begin{array}{c} 
n \\
n \end{array} \right]_{(q,h)} .
\end{eqnarray*}
This completes the Proof. \\  
~\\
Moreover, the $(q,h)$--binomial coefficients obey to the following   
properties $1<k<n$ :
\begin{eqnarray}
\left[ \begin{array}{c} 
n + 1 \\
k \end{array} \right]_{(q,h)} & = & 
q^k ~\left[ \begin{array}{c} 
n \\
k \end{array} \right]_{(q,h)}~ + 
~ ( 1 + h~[k-1]_q ) \left[ \begin{array}{c} 
n \\
k - 1 \end{array} \right]_{(q,h)} 
\end{eqnarray}
and 
\begin{eqnarray}
\left[ \begin{array}{c} 
n + 1 \\
k + 1 \end{array} \right]_{(q,h)} & = & ( 1 + h~[k]_q )   
\frac{[n+1]_q}{[k+1]_q}  \left[ 
\begin{array}{c}
n \\
k \end{array} \right]_{(q,h)}.
\end{eqnarray}
In fact, these properties follow from the well--known relations of the 
$q$--binomial coefficients : 
\begin{eqnarray*}
\left[ \begin{array}{c}
n + 1 \\
k \end{array} \right]_q & = & q^k~\left[ \begin{array}{c} 
n \\
k \end{array} \right]_q + \left[ \begin{array}{c}
n \\
k - 1 \end{array} \right]_q
\end{eqnarray*}
and 
\begin{eqnarray*}  
\left[ \begin{array}{c} 
n + 1 \\
k \end{array} \right]_q & = & 
\frac{[n + 1]_q}{[k]_q}~\left[ \begin{array}{c}
n \\
k - 1 \end{array} \right]_q
\end{eqnarray*}
upon using $(a)_{[k]} = (a + {[k -1]_q} ) (a)_{[k-1]}$. \\ 
~\\
Now, we make the following remarks. For $h=0$ this is just the $q$--
binomial formula as it should be. \\ 
For $q=1$, it reduces to the $h$--analogue Newton's binomial 
formula (3) and (4) and for $(q=1, h=0)$ the usual one is recovered. \\
~\\
To conclude, we have obtained a more general Newton's binomial formula in 
$(q,h)$--deformed quantum plane which reduces to the known one at some 
limits. This will lead therefore to a more generalized analysis called 
$(q,h)$--analysis.  
\section*{acknowlegment}
I'd like to thank the DAAD for its financial support and the referee for his 
remarks.    

~\\

\end{document}